\documentstyle[preprint,aps]{revtex}
\draft
\begin{document}
\begin{title}
{Griffiths singularity in the two dimensional random
bond disordered Ising ferromagnet}
\end{title}
\author{Jae-Kwon Kim}
\date{\today}
\address
{Department of Physics, University of California,
Los Angeles, CA 90024}
\maketitle

\begin{abstract}
For the two dimensional
random bond disordered Ising ferromagnet, we measured bulk data
of the magnetic susceptibility ($\chi$) and correlation length
($\xi$) up to $\xi \simeq 536$, with the use of a novel
finite size scaling Monte Carlo technique.
Our data are exclusively consistent with normal power-law
critical behaviors
with only one singular point at criticality, disproving
the existence of Griffiths  singularity even in an extremely
deep scaling region.
The critical exponents of $\chi$ and $\xi$ increase continuously
with the strength of disorder.
\end{abstract}

	The critical behavior of the  two-dimensional (2D)
	randomly disordered Ising ferromagnet is an outstanding
	problem in both condensed matter and statistical physics.
	By the random disorder is  meant either a random site
	dilution  or random- valued positive coupling. The effect of
	the fluctuations of the quenched random disorder on
	the critical behavior is of main interest.
	McCoy and Wu\cite{MCO} made the first successful attempt
	for such a model, and
	proved that the specific heat ($C_{v}$) is non-divergent in the 2D
	Ising system with one-directional and correlated random
	bond disorder.
	According to McCoy and Wu, the divergence of $C_{v}$
	is caused by the coherence of all the bonds acting together,
	so destroying the coherence by introducing one or two
	directional bond disorders makes $C_{v}$ finite.

	Many authors, however, regarded the non-diverging behavior of
	$C_{v}$ in the McCoy -Wu model
	as a characteristic of the one-dimensional correlated disorder.
	Especially, some authors,
	based on their observation that
	the continuum limit of the 2D random bond disordered Ising model
	is a certain type of Gross-Neveu model,
	predicted $\eta=1/4$ \cite{SHA}
	along with double-logarithmically diverging critical behavior
	of $C_{v}$\cite{SHA}\cite{DOT}. Namely, for
	weakly disordered 2D random bond disordered  Ising (RBDI)
	ferromagnet it was predicted:
	\begin{eqnarray}
	\xi &\sim & t^{-\nu} [1 + C\ln(1/t)]^{\tilde{\nu}},
	~~\nu=1,~ \tilde{\nu}=1/2 \label{eq:cor} \\
	\chi &\sim & \xi^{2-\eta},~~\eta=1/4  \label{eq:suc}\\
	C_{v} &\sim & t^{-\alpha} \ln[1 + C\ln(1/t)] + C^{\prime}, ~~
	\label{eq:sp}
	\alpha=0
	\end{eqnarray}
	with $t$ representing
	the reduced (inverse) temperature.
	A heuristic criterion initially given by Harris\cite{HAR}
	is reflected in these expressions;
	that is, a random quenched  disorder
	is relevant only if $\alpha$ of the pure system is positive,
	and only in deep scaling region close to criticality.

	Now the theoretical prediction of $\eta=1/4$
	has been confirmed by a different theoretical
	approach\cite{JUG}
	along with by numerous numerical studies\cite{WANG}
	\cite{QUE}\cite{PAT}. The predicted behavior of
	$C_{v}$, however, contradicts some other analytical results
	\cite{TIMO}\cite{ZIEG} and recent extensive numerical work
	on the 2D randomly site diluted Ising system (RSDI)\cite{PAT}.
	Various numerical methods\cite{PAT}\cite{RAP}\cite{KUHN}\cite{FAH}
	report that the presence of strong random site dilutions
	actually changes the universality class; nevertheless,
	a conclusive numerical evidence lacks in the 2D random
	bond disordered Ising (RBDI) case where the effect of
	disorder is much weaker than in RSDI, albeit some numerical
	results supporting the scenario of the logarithmic
	correction\cite{WANG}\cite{DER}\cite{TALA}.

	Another important issue on the 2D RSDI  was first
	addressed by Griffiths\cite{GRIF}, who argued that
	magnetization as a function of external field ($H$)
	becomes non-analytic at $H=0$ below a temperature nowadays
	referred to a Griffiths temperature ($T_{G}$).
	Being identified as the critical
	temperature of the fictitious pure system with the
	largest value of the coupling allowed in the random system,
	$T_{G}$ is independent of the concentration of random dilution,
	and lies {\it above} the the critical temperature ($T_{c}$).
	Griffiths phase which refers to a range of temperature
	where Griffiths singularity takes place, i.e.,
	$T_{c} \le T \le T_{G}$, however, prolongs with the
	degree of dilution.

	Griffiths singularity has been elaborated and
	extended by many authors\cite{BRAY}\cite{CAR}; especially,
	it is claimed that in Griffiths phase free energy is
	singular as a function of temperature as well at $H=0$\cite{NIEU}.
	Ziegler's\cite{ZIEG} argument contrasts among others, however,
	in that only two singular points
	(not a Griffiths {\it phase}) exist in the 2D
	randomly disordered Ising ferromagnet
	at which both $\chi$ and $\xi$ diverge but $C_{v}$ does not.

	Griffiths singularity is believed to be generic in any disordered
	system, including spin glasses\cite{CHA} and random fields\cite{DOT2}.
        However, few numerical attempts to prove or to disprove
	the presence of Griffiths singularity were made, mainly
	because it is not clear how to treat Griffiths singularity
	in the context of either the standard finite size scaling (FSS) or
	Monte Carlo renormalization group method. In fact,
	the non-existence of Griffiths singularity is prerequisite
	for a reliable application of the Monte Carlo renormalization
	group method to a randomly disordered system \cite{FAH}.

	In this Letter, we attempt to clarify the unresolved
	issues  of the 2D RBDI ferromagnet by the MC measurements
	of the bulk data (thermodynamic data) up to an extraordinarily
	deep scaling region\cite{COM1}.
	In general, with a knowledge of
	bulk data (thermodynamic data) in a deep scaling region
	all the necessary informations
	regarding a critical behavior can be gained
	in the most straightforward way.
	If Griffiths phase indeed exists, one expects to
	observe either a divergence or a discontinuity in  the
	behaviors of $\chi (t)$ and $\xi (t)$ at temperatures
	other than criticality. The broad range of the
	bulk data is also necessary to test against the claims of
	some authors \cite{TALA}\cite{SEL}\cite{HEU} that
	only some {\it effective} $\gamma$ and $\nu$ increase
	with the strength of disorder.

	Practically, a physical quantity measured
	on a finite system with a linear size of L at
	a $t$, say $P_{L}(t)$,
	converges to its thermodynamic value (bulk value),
	$P_{\infty}(t)$, under the {\it thermodynamic}
	condition $\xi_{L}/L \ge r$.
	The value of $r$ which is independent of temperature
	is approximately 6  for the
	2D pure Ising ferromagnet with periodic boundary condition
	\cite{PAT1},
	but increases with the strength of random disorder\cite{PAT}.
	Consequently, with the use of traditional Monte Carlo methods
	it is prohibitively difficult to measure proper thermodynamic
	values at temperatures sufficiently close to a critical point.

Our evaluations of bulk values are based
on a novel technique recently developed by Kim\cite{KIME}.
Using this technique, the bulk data can be extrapolated with
MC data obtained on much smaller lattice.
The technique is based on the fundamental formula of finite
size scaling\cite{KIME}.
Namely, for a multiplicative renormalizable quantity $P$
\begin{equation}
P_{L}(t)=P_{\infty}(t) q_{P}(x),~~~x=\xi_{L}/L, \label{eq:fss}
\end{equation}
where $P_{L}$ denotes $P$ defined on a finite lattice of linear size L.
Notice that here we use a scaling variable  different
from the traditional one, $\xi_{\infty}/L$,
and that $q_{P}(x)$ represents a universality class.
Because of a correction to FSS mainly due to
certain irrelevant operators, Eq.(\ref{eq:fss}) is valid
only for $L \ge L_{min}$, where $L_{min} \simeq 20$
for most models without crossover in the critical behavior.

The outline of the technique is as follows:
The functional value of $q_{P}(x)$
is easily determined numerically for some discrete values
of $x$, at a temperature
where $P_{\infty}$ including $\xi_{\infty}$ is already known.
Since $q_{P}(x)$ has no explicit
temperature dependence, for a given $x^{\prime}$,
the value of $q_{P}(x^{\prime})$ at any
other temperature can be interpolated
based on these known values of $q_{P}(x)$.
Once $q_{P}(x^{\prime})$ is determined,
$P_{\infty}$ is easily computed from Eq.(\ref{eq:fss})
by the measurement of $P_{L}$ with the value of L
corresponding to the $x^{\prime}$. An interpolation at
larger $x^{\prime}$ makes this technique more efficient.
At the price of using modestly small values of L
this technique requires very precise measurements;
nevertheless, the cost of computer
resources for the required precision is much less
than simulations of huge lattice system.

	Our Hamiltonian is defined as
	\begin{equation}
	H=- \sum_{<ij>} J_{ij} S_{i}S_{j},~~~S_{i}=\pm 1,
	\end{equation}
	where the sum  is over all the nearest neigbors of
	lattice.
	$J_{ij}$ is defined to be positive, e.g., taking
	either a positive valued $J$ or $J^{\prime}$ randomly with
	probability $p$ and $1-p$ respectively.
	For $p=1/2$, the system is self-dual\cite{FISC}
	with the self-dual point given by
	\begin{equation}
	\tanh (J)= \exp (-2J^{\prime})
	\end{equation}
	A self-dual point equals the critical point of
	a system, provided the system has only one critical point.
	We fix $J=1$ and $p=1/2$ without any loss of generality,
	and consider three different values of $J^{\prime}$, i.e.,
	$J^{\prime}= 0.9$, 0.25, and 0.1.
	Accordingly, the  inverse Griffiths temperature, $\beta_{G}$,
	is equivalent to the critical point of pure Ising system, i.e.
	$\beta_{G}=\ln(\sqrt 2+1)/2$, with the corresponding inverse
	self-dual points (critical points)
	given by $\beta_{c}=0.4642819\ldots$,
	$0.80705185\ldots$, and $1.10389523\ldots$  for
	$J^{\prime}=0.9$, 0.25, and 0.1 respectively.

	Our raw-data for each $J^{\prime}$ are obtained by choosing
	a realization of random $J^{\prime}$, then running
	Monte Carlo simulations
	in Wolff's one cluster algorithm with periodic
	boundary conditions; for each realization,
	measurements were taken over 10 000 configurations
	each of which was separated by 2-8 one cluster updatings
	according to auto-correlation times.
	The procedure is then repeated for different
	realizations of $J^{\prime}$.
	The average over all the different realizations converges
	as the numbers of the random realization
	increase; basically this mean value of
	a physical quantity is something physically interesting.
	To achieve the necessary precision for our technique,
	the numbers of the different realizations we used are
	approximately 20 - 40, 150 - 250, and 300 - 600
	for $J^{\prime}=0.9$, 0.25, and 0.1 respectively; yet, in general,
	the fluctuation among different realizations of the random
	disorder is more significant than the statistical error
	for a given realization. Consequently, we ignore here
	the latter in our calculation of the statistical  error
	of a mean value.
	The largest value of L we used to extract our bulk values
	is just 240. The total computing time spent for this
	work amounts to  more than 600 CPU hours in unit of
	CRAY YMP832 Supercomputer.

	For $J^{\prime}=0.9$, 0.25, and 0.1 respectively,
	the ranges of our correlation lengths thus evaluated
	are over $5.7(1) \le \xi \le 536.0(8.1)$ ([5.7(1), 536.0(8.1)]),
	[5.8(1), 429.2(15.5)], and [5.0(2), 403.4(22.3)],
	corresponding to the range of $\beta$ (inverse temperature)
	over $0.42 \le \beta \le 0.4638$
	([0.42, 0.4638]), [0.70, 0.805], and [0.87, 1.097].
	We advertently choose the range of $\beta$ such
	that $\xi(\beta) \ge 5$, so that the effect of
	certain nonconfluent correction to scaling can be
	unimportant in the fits. (See below.)
	Our bulk data are summerized in Figs.(1) and (2),
	which depict $\ln \xi(t)$ and $\ln \chi(t)$ as a function of
	$|\ln t|$ respectively.  Note that the slopes in Figs.(1) and
	(2) respectively represent $\nu$ and $\gamma$.

	The data show no sign of singular behavior such as
	a divergence or a discontinuity at the temperatures
	in the Griffiths phase, up to $t \simeq 1.1 \times 10^{-3}$.
	The almost perfect straight lines in such broad
	ranges of the bulk data, showing that both $\gamma$ and
	$\nu$ do not change with $t$, clearly prove
	that they do not represent the effective values of the
	critical exponents but the asymptotic ones\cite{COM2}.
	Thus, the data for each $J^{\prime}$ are exclusively
	consistent with normal
	power-law singularities having one and only one singular
	point at criticality.
	With the conspicuous differences in the slope,
	the  values of $\nu$ and $\gamma$ obviously increase
	with decreasing $J^{\prime}$.

	Fixing the critical points at the self-dual points in the
	$\chi^{2}$ fits and assuming a pure power-law type
	critical behavior,
	we obtain $\nu=1.00(0)$ and $\gamma=1.75(1)$ ([1.00(1), 1.75(1)]),
	[1.09(1), 1.90(2)], and [1.23(2), 2.13(3)] for
	$J^{\prime}=0.9$, 0.25, and  0.1 respectively.
	Notice that $\eta=2-\gamma/\nu$ remains a constant i.e.,
	$\eta=1/4$ within the statistical errors,
	irrespective of the values of $J^{\prime}$.
	Assuming a scaling function with a nonconfluent
	correction term, e.g., $\xi(t) \sim t^{-\nu}(1+a t),$
	yields no significant change in the estimate of the critical
	exponent, e.g., $\nu=1.075(7)$ for $J^{\prime}=0.25$.

	For $J^{\prime}=0.9$ the estimated values of
	$\nu$ and $\gamma$ are virtually the same
	as those for the pure system.
	Nevertheless, we do not take this as an evidence that
	the critical exponents are strictly the same in the two
	systems,
	due to the general tendency of increasing $\nu$ and $\gamma$
	with decreasing $J^{\prime}$.
	Rather, it appears that the values of $\nu$ and $\gamma$ increase
	so mildly for weakly disordered system ($J^{\prime} \simeq 1$)
	that they are extremely hard to be distangled from those
	of the pure system.

	Distinguishing a tiny
	increase of $\nu$ from a multiplicative logarithmic correction
	by comparing $\chi^{2}/N_{DF}$ values
	of the two fits is practically very difficult;
	accordingly, for $J^{\prime}=0.9$ and 0.25
	the data fit to the Eqs.(1) and (2) as well as to the pure
	power-laws.
	For $J^{\prime}=0.1$, however, the $\xi$ ($\chi$)
	data fit to a pure power-law much better than to Eq.(1) (Eq.(2)):
	The values of $\chi^{2}/N_{DF}$ for the two fits of the $\xi$
	data are 4.9 and 0.7, respectively to Eq.(1) and to
	the pure power-law type.  Thus, for a strong RBDI case at least,
	the scenario of the logarithmic correction is clearly  unfavored
	by the data.
	Although for $J^{\prime}=0.9$ and 0.25
	we are unable to determine the correct type of
	the critical behavior through the $\chi^{2}$ fit,
	the universal scaling functions, $q_{P}(x)$ in Eq.(\ref{eq:fss}),
	for the two values of $J^{\prime}$ are  definitely
	different from each other,
	as shown in Fig.(3), indicating that they indeed
	belong to different universality classes.

	According to the standard theory of FSS,
	the value of Binder's cumulant ratio\cite{BIN} at
	criticality ($U_{L}(t=0)$) for a fixed geometry
        is another indicator of a universality class\cite{PRIV}.
	Our data of $U_{L}(t=0)$ in the range $20 \le L \le 100$
	are tabulated in Tab.(1), showing
	that $U_{L}(t=0)$ for each $J^{\prime}$ is
	indeed independent of L within the statistical errors,
	and that $U_{L}(t=0)$ increases with decreasing $J^{\prime}$.
	Note that $U_{L}(t=0)$ for $J^{\prime}=0.9$ are virtually
	the same as for the pure system, i.e.,
	$U_{L}(t=0) =1.832(1)$\cite{PAT1}, as $\nu$ and $\gamma$ are
	for the same $J^{\prime}$.
	Indeed, the behavior of $U_{L}(t=0)$ as a function
	of $J^{\prime}$ confirms our claim that $\nu$ and $\gamma$
	increase continuously with the strength of the disorder.

	It is unlikely that the increase of $U_{L}(t=0)$ is owing to a
	logarithmic correction\cite{DER} for the following two reasons:
        (i) Notice that the value of $\tilde{\nu}$  in Eq.(1)
	is fixed regardless of the strength of the random disorder, i.e.,
	only the non-universal constant $C$ varies with it. Therefore,
	the continuous variance of $U_{L}(t=0)$ with respect to
	$J^{\prime}$ is improbable
	in the context of the scenario of the logarithmic correction.
	(ii) More crucially, it is observed\cite{KIM3} that
	$U_{L}(t=0)$ changes very mildly along the critical line
	of the Ashkin-Teller model
	along which $\nu$ varies continuously.

	Some remarks are in order: (i)
	Because of the mild variation of the critical
	exponents, along with the hyperscaling
	relation, $\alpha=2-D\nu$, we expect that it would be very
	difficult to observe a finite peak of $C_{v}$
	in  weakly disordered case.
	Notice, however, that this is the case
	even in the 3D (pure) XY model\cite{SCH}, where $\alpha <0$.
	Moreover, in this model the specific heat at criticality
	increase with L monotonically\cite{SCH},
	so that it can be fitted to a double logarithmic function.
	This is a clear demonstration that a mild increase
	of the specific heat at criticality, for some finite values of L,
	cannot be an evidence for its divergence at criticality\cite{WANG}.
	(ii) According to Ziegler\cite{ZIE2}, his singular point manifests
	itself at the length scale $L\simeq 900$ for $J^{\prime}=0.25$.
	With our largest $\xi \simeq 429$ and with our observation that
	$r \simeq 8$ for $J^{\prime}=0.25$, our largest length scale
	amounts to $L \simeq 429 \times 8 \simeq 3430$, which is
	far beyond his threshold length scale.
	(iii) In the context of the renormalization group approach,
	the continuous variation of a critical exponent
	implies the existence of a line of the fixed points,
	as in the 2D Ashkin- Teller model.
	We conjecture that even in the 3D randomly disordered
	Ising ferromagnet the
	critical exponents would vary continuously, instead of
	the presence of an additional fixed point
	traditionally advocated\cite{CHAY}.

	To conclude:
	We have studied the critical behavior of the
	2D RBDI ferromagnet with
	unprecedented extensiveness and precision, so that
	some conclusive results of its critical behavior
	are obtained.
	Our data unambiguously show that both $\nu$ and $\gamma$
	increase continuously with the strength of bond disorder.
	Any scenarios of Griffiths singularity
	are not respected by the data.

	The author would like to thank  Hyunggyu Park,
	S.L.A. de Queiroz, Walter Selke, and Klaus Ziegler
	for the exchanges,
	Jang Jin Chae, Jaeshin Lee, and Ghi-Ryang Shin for their
	continuous support for computing.

\begin{table}
\caption{Binder's cumulant ratio at the self-dual points, for
	 the three values of $J^{\prime}$.
	 Note that $U_{L}(t=0)$ for each $J^{\prime}$ does not
	 vary with L within the statistical errors,
	 thus showing that each self-dual
	 point is indeed the critical point. Also,
	 it is clear that $U_{L}(t=0)$ increases with
	 decreasing $J^{\prime}$,
	 although for $J^{\prime}=0.9$ it is virtually the same
	 as in the pure system.}
\begin{tabular}{cccc}
       &$J^{\prime}=0.90$    &$J^{\prime}=0.25$    &$J^{\prime}=0.10$ \\
	\hline
L=20   &1.834(1)              &1.850(3)             &1.862(3)   \\
L=40   &1.833(3)              &1.847(3)		    &1.858(3)   \\
L=60   &1.832(1)	      &1.851(4)		    &1.854(4)    \\
L=80   &1.833(1)	      &1.849(3)		    &1.862(4)   \\
L=100  &1.832(2)              &1.855(6)		    &1.858(4)   \\
\end{tabular}
\end{table}


\begin{references}
\bibitem{MCO} B.M. McCoy and T.T. Wu, Phys. Rev. {\bf 176} 631 (1968);
		{\it ibid},
		{\it The two dimensional Ising Model} (Harvard Univ.
		 Press, Cambridge, 1973)
\bibitem{SHA} B.N. Shalaev, Sov. Phys. Solid State {\bf 26}, 1811 (1984);
	   {\it ibid}, Phys, Rep. {\bf 237}, 129 (1994);
	   R. Shankar, Phys. Rev. Lett. {\bf 58}, 2466 (1987);
	   A.W.W. Ludwig, {\it ibid} {\bf 61}, 2388 (1988)
\bibitem{DOT} V. Dotsenko and V. Dotsenko, Adv. Phys. {\bf 32}, 129 (1983)
\bibitem{HAR} A.B. Harris, J. Phys. C {\bf 7}, 1671 (1974)
\bibitem{JUG}  G. Jug, Phys. Rev. Lett. {\bf 55}, 1343 (1985)
\bibitem{WANG} J.-S. Wang, W. Selke, V.S. Dotsenko and V.B. Andreichenko,
		Physica A {\bf 164}, 221 (1990);
	    V.B. Andreichenko, V.S. Dotsenko, W. Selke and J.-S. Wang,
	    Nucl. Phys. B {\bf 344}, 531 (1990)
\bibitem{QUE} S.L.A. de Queiroz and R.B. Stinchcombe, Phys. Rev. B {\bf 46},
	   6635 (1992); {\it ibid}, {\bf 50}, 9976 (1994)
\bibitem{PAT} J.-K. Kim and A. Patrascioiu, Phys. Rev. Lett. {\bf 72},
	   2785 (1994);{\it ibid}, Phys. Rev. B {\bf 49}, 15764  (1994)
\bibitem{TIMO} P.-N. Timonin, Sov. Phys. -JEPT {\bf 95}, 893 (1989)
\bibitem{ZIEG} K. Ziegler, Nucl. Phys. B {\bf 344}, 499 (1990); D. Braak
	    and K. Ziegler, Z. Phys. B {\bf 89}, 361 (1992)
\bibitem{RAP} D.C. Rapaport, J. Phys. C {\bf 5}, 2813 (1972)
\bibitem{KUHN} R. K\"{u}hn, Phys. Rev. Lett. {\bf 73}, 2268 (1994)
\bibitem{FAH} M. Faenhle, T. Holey, and J. Eckert, J. Magn. Magn. Mater.
		{\bf 195}, 104 (1992)
\bibitem{DER} B. Derrida, B.W. Southern, and D. Stauffer,
	       J. Phys. (Paris), {\bf 48}, 335 (1987)
\bibitem{TALA}  A.L. Talapov and L.N. Shchur, hep-lat/9404011 (to be published)
\bibitem{GRIF} R.B. Griffiths, Phys. Rev. Lett. {\bf 23}, 17 (1969)
\bibitem{BRAY} A.J. Bray and D. Huifang, Phys. Rev. B {\bf 40}, 6980 (1989)
\bibitem{CAR} A.W.W Ludwig and J.L. Cardy, Nucl. Phys. B {\bf 285} [FS19],
	       687 (1987)
\bibitem{NIEU} Th. M. Nieuwenhuizen, Phys. Rev. Lett. {\bf 63}, 1760 (1989)
\bibitem{CHA} For the 2D Ising spin glass, however, a study by the
	      high- temperature series finds  no evidence for
	      Griffiths phase. R.R.P. Singh and S. Charkravarty,
	      Phys. Rev. B {\bf 36}, 559 (1987)
\bibitem{DOT2} See, for example, V. Dotsenko, J. Phys. A {\bf 27},
	       3397 (1994), and references therein.
\bibitem{COM1} Some previous attempt\cite{TALA}\cite{HEU}
	      to measure thermodynamic values of $\chi$
	      reported a dissenting conclusion with this work.
	      Without measuring  correlation length, however, it appears
	      that the data at temperatures closer to criticality
	      are flawed by the finite size effect.
\bibitem{SEL} W. Selke, Phys. Rev. Lett. {\bf 73}, 3487 (1994)
\bibitem{HEU} H.-O. Heuer, Phys. Rev. B {\bf 45}, 5691 (1992)
\bibitem{PAT1} J.-K. Kim and A. Patrascioiu, Phys. Rev. D {\bf 47},
	       2588 (1993)
\bibitem{KIME} J.-K. Kim, Europhys. Lett. {\bf 28}, 211 (1994);
	    {\it ibid}, Nucl. Phys. B [Proc. Suppl.] {\bf 34}, 702 (1994)
\bibitem{FISC} R. Fisch, J. Stat. Phys. {\bf 18}, 111 (1978)
\bibitem{COM2} In general, an effective value of
	a critical exponent  that is different from the asymptotic
	one manifests itself when the data at temperatures
	such that $\xi(t) << 5$ are taken into account,
	as can be easily checked with the exact formula of $\xi(t)$
	for the 2D pure Ising model. Also, the value
	of an effective critical exponent varies with temperature.
\bibitem{BIN} K. Binder, Phys. Rev. Lett. {\bf 47}, 693 (1981)
\bibitem{PRIV} See, for example, V. Privmann in {\it Finite Size Scaling
	 and Numerical Simulations of Statistical Systems}, edited
	 by V. Privman (World Scientific, 1990)
\bibitem{KIM3} J.-K. Kim (unpublished)
\bibitem{SCH} N. Schultka and E. Manousakis, preprint,
	      cond-mat/9502062 (1995)
\bibitem{ZIE2} K. Ziegler, private communication
\bibitem{CHAY} J.T. Chayes, L. Chayes, D.S. Fisher, and T. Spencer,
		Phys. Rev. Lett. {\bf 57}, 2999 (1986)


{\bf Figure Captions} \\
Fig.(1): $\ln\xi$ versus $|\ln t|$.
	The dotted lines represent the results of
	the best $\chi^{2}$ fits assuming
	pure power-law type singularity. The values of the slope
        , which correspond to the values of $\nu$, are 1.00, 1.09,
	and 1.23 respectively for $J^{\prime}=0.90$, 0.25, and 0.10.
	The data for the pure Ising model ($J^{\prime}=1$) are taken
	from the well- known exact formula,
	over the range $4.75 \le \xi \le 425.9$.
	Notice  that they are precisely on the line of $\nu=1$,
	showing that the correct scaling region already sets in
	at a temperature where $\xi \simeq 5$. \\
Fig.(2): $\ln\chi$ versus $|\ln t|$. Here the slope of each line
	  corresponds to the value of $\gamma$, for each $J^{\prime}$. \\
Fig.(3): $q_{P}$ versus $x$. The upper two {\it curves} are for $P=\xi$,
	 while the lower two are for $P=\chi$. The data on each curve actually
	 represent  data-collapse from different temperatures. (The detailed
	 pictures will be presented elsewhere.)
	 For each P, it is obvious that each $J^{\prime}$ characterizes
 	 a different curve, i.e., a different universality class.
\end{references}
\end{document}